\documentstyle[epsf,eqsecnum,aps,floats]{revtex}
\parskip \baselineskip
\begin{document}
\def\theequation{\arabic{equation}}%
\newcommand{\rstev}{\mbox{$\rs = \T{1.8}$}}
\newcommand{\XX}{\mbox{$\, \times \,$}}
\newcommand{\AP}{\mbox{${\rm \bar{p}}$}}
\newcommand{\SU}{\mbox{$S$}}
\newcommand{\SPt}{\mbox{$<\! |S|^2 \!>$}}
\newcommand{\ET}{\mbox{$E_{T}$}}
\newcommand{\PT}{\mbox{$p_{t}$}}
\newcommand{\DP}{\mbox{$\Delta\phi$}}
\newcommand{\DR}{\mbox{$\Delta R$}}
\newcommand{\DE}{\mbox{$\Delta\eta$}}
\newcommand{\DEP}{\mbox{$\Delta\eta_{c}$}}
\newcommand{\DEC}{\mbox{$\Delta\eta_{c}$}}
\newcommand{\SP}{\mbox{$S(\DEP)$}}
\newcommand{\PH}{\mbox{$\phi$}}
\newcommand{\EA}{\mbox{$\eta$}}
\newcommand{\EAJ}{\mbox{\EA(jet)}}
\newcommand{\AEA}{\mbox{$|\eta|$}}
\newcommand{\Ge}[1]{\mbox{#1 GeV}}
\newcommand{\T}[1]{\mbox{#1 TeV}}
\newcommand{\D}[1]{\mbox{$#1^{\circ}$}}
\newcommand{\x}{\cdot}
\newcommand{\ra}{\rightarrow}
\newcommand{\mb}{\mbox{mb}}
\newcommand{\nb}{\mbox{nb}}
\newcommand{\ipb}{\mbox{${\rm pb}^{-1}$}}
\newcommand{\inb}{\mbox{${\rm nb}^{-1}$}}
\newcommand{\rs}{\mbox{$\sqrt{s}$}}
\newcommand{\fdel}{\mbox{$f(\DEP)$}}
\newcommand{\fdele}{\mbox{$f(\DEP)^{exp}$}}
\newcommand{\fgap}{\mbox{$f(\DEP\! > \!3)$}}
\newcommand{\fgape}{\mbox{$f(\DEP\! > \!3)^{exp}$}}
\newcommand{\fpyt}{\mbox{$f(\DEP\!>\!2)$}}
\newcommand{\delth}{\mbox{$\DEP\! > \!3$}}
\newcommand{\uplim}{\mbox{$1.1\!\times\!10^{-2}$}}
\newcommand{\sigew}{\mbox{$\sigma_{\rm EW}$}}
\newcommand{\sigsi}{\mbox{$\sigma_{\rm singlet}$}}
\newcommand{\sigr}{\mbox{$\sigsi/\sigma$}}
\newcommand{\sigrew}{\mbox{$\sigew/\sigma$}}
\newcommand{\ncal}{\mbox{$n_{\rm cal}$}}
\newcommand{\ntrk}{\mbox{$n_{\rm trk}$}}

\def\simge
{\mathrel{\rlap{\raise 0.53ex \hbox{$>$}}{\lower 0.53ex \hbox{$\sim$}}}}

\def\simle
{\mathrel{\rlap{\raise 0.4ex \hbox{$<$}}{\lower 0.72ex \hbox{$\sim$}}}}

%
\def\sigtot{$\sigma_{\rm tot}$}         
\def\sigtop{$\sigma_{t \overline{t}}$}  
\def\pbarp{$\overline{p}p $}            
\def\ppbar{$p\overline{p} $}            
\def\qqbar{$q\overline{q}$}             
\def\ttbar{$t\overline{t}$}             
\def\bbbar{$b\overline{b}$}             
\def\D0{D\O}                            
\def\ipb{pb$^{-1}$}                     
\def\pt{p_T}                            
\def\ptg{p_T^\gamma}                    
\def\et{E_T}                            
\def\etg{E_T^\gamma}                    
\def\htran{$H_T$}                       
\def\gevcc{{\rm GeV}/c$^2$}                   
\def\gevc{{\rm GeV}/c}                  
\def\gev{\rm GeV}                       
\def\tev{\rm TeV}                       
\def\njet{$N_{\rm jet}$}                
\def\aplan{$\cal{A}$}                   
\def\lum{$\cal{L}$}                     
\def\iso{$\cal{I}$}                     
\def\remu{${\cal{R}}_{e\mu}$}           
\def\rmu{$\Delta\cal{R}_{\mu}$}         
\def\pbar{$\overline{p}$}               
\def\tbar{$\overline{t}$}               
\def\bbar{$\overline{b}$}               
\def\lumint{$\int {\cal{L}} dt$}        
\def\lumunits{cm$^{-2}$s$^{-1}$}        
\def\etal{{\sl et al.}}                 
\def\vs{{\sl vs.}}                      
\def\sinthw{sin$^2 \theta_W$}           
\def\mt{$m_t$}                          
\def\mb{$m_b$}                          
\def\mw{$M_W$}                          
\def\mz{$M_Z$}                          
\def\pizero{$\pi^0$}                    
\def\jpsi{$J/\psi$}                     
\def\wino{$\widetilde W$}               
\def\zino{$\widetilde Z$}               
\def\squark{$\widetilde q$}             
\def\gluino{$\widetilde g$}             
\def\alphas{$\alpha_{\scriptscriptstyle S}$}                
\def\alphaem{$\alpha_{\scriptscriptstyle{\rm EM}}$}         
\def\epm{$e^+e^-$}                      
\def\deg{$^\circ$}                      
\def\met{\mbox{${\hbox{$E$\kern-0.6em\lower-.1ex\hbox{/}}}_T$ }} 
\newcommand{\NC}{{\em Nuovo Cimento\/} }
\newcommand{\NIM}{{\em Nucl. Instr. Meth.} }
\newcommand{\NP}{{\em Nucl. Phys.} }
\newcommand{\PL}{{\em Phys. Lett.} }
\newcommand{\PR}{{\em Phys. Rev.} }
\newcommand{\PRL}{{\em Phys. Rev. Lett.} }
\newcommand{\RMP}{{\em Rev. Mod. Phys.} }
\newcommand{\ZP}{{\em Zeit. Phys.} }
\def\err#1#2#3 {{\it Erratum} {\bf#1},{\ #2} (19#3)}
\def\ib#1#2#3 {{\it ibid.} {\bf#1},{\ #2} (19#3)}
\def\nc#1#2#3 {Nuovo Cim. {\bf#1} ,#2(19#3)}
\def\nim#1#2#3 {Nucl. Instr. Meth. {\bf#1},{\ #2} (19#3)}
\def\np#1#2#3 {Nucl. Phys. {\bf#1},{\ #2} (19#3)}
\def\pl#1#2#3 {Phys. Lett. {\bf#1},{\ #2} (19#3)}
\def\prev#1#2#3 {Phys. Rev. {\bf#1},{\ #2} (19#3)}
\def\prl#1#2#3 {Phys. Rev. Lett. {\bf#1},{\ #2} (19#3)}
\def\rmp#1#2#3 {Rev. Mod. Phys. {\bf#1},{\ #2} (19#3)}
\def\zp#1#2#3 {Zeit. Phys. {\bf#1},{\ #2} (19#3)}

\title{
\begin{flushright}
Fermilab-Pub-96/072-E\\
\end{flushright}
\vspace{-0.2cm}
The Isolated Photon Cross Section in the 
Central and Forward Rapidity Regions\\
in $p \overline p$ Collisions at \rstev}

%
\author{                                                                        
S.~Abachi,$^{14}$                                                               
B.~Abbott,$^{28}$                                                               
M.~Abolins,$^{25}$                                                              
B.S.~Acharya,$^{44}$                                                            
I.~Adam,$^{12}$                                                                 
D.L.~Adams,$^{37}$                                                              
M.~Adams,$^{17}$                                                                
S.~Ahn,$^{14}$                                                                  
H.~Aihara,$^{22}$                                                               
J.~Alitti,$^{40}$
% ...
\\                                                                              
\vskip 0.50cm                                                                   
\centerline{(D\O\ Collaboration)}                                               
\vskip 0.50cm                                                                   
}                                                                               
\address{                                                                       
\centerline{$^{1}$Universidad de los Andes, Bogot\'{a}, Colombia}               
\centerline{$^{2}$University of Arizona, Tucson, Arizona 85721}                 
\centerline{$^{3}$Boston University, Boston, Massachusetts 02215}               
\centerline{$^{4}$Brookhaven National Laboratory, Upton, New York 11973}        
\centerline{$^{5}$Brown University, Providence, Rhode Island 02912}             
\centerline{$^{6}$Universidad de Buenos Aires, Buenos Aires, Argentina}         
\centerline{$^{7}$University of California, Davis, California 95616}            
\centerline{$^{8}$University of California, Irvine, California 92717}           
\centerline{$^{9}$University of California, Riverside, California 92521}        
\centerline{$^{10}$LAFEX, Centro Brasileiro de Pesquisas F{\'\i}sicas,          
                  Rio de Janeiro, Brazil}                                       
\centerline{$^{11}$CINVESTAV, Mexico City, Mexico}                              
\centerline{$^{12}$Columbia University, New York, New York 10027}               
\centerline{$^{13}$Delhi University, Delhi, India 110007}                       
\centerline{$^{14}$Fermi National Accelerator Laboratory, Batavia,              
                   Illinois 60510}                                              
\centerline{$^{15}$Florida State University, Tallahassee, Florida 32306}        
\centerline{$^{16}$University of Hawaii, Honolulu, Hawaii 96822}                
\centerline{$^{17}$University of Illinois at Chicago, Chicago, Illinois 60607}  
\centerline{$^{18}$Indiana University, Bloomington, Indiana 47405}              
\centerline{$^{19}$Iowa State University, Ames, Iowa 50011}                     
\centerline{$^{20}$Korea University, Seoul, Korea}                              
\centerline{$^{21}$Kyungsung University, Pusan, Korea}                          
\centerline{$^{22}$Lawrence Berkeley National Laboratory and University of      
                   California, Berkeley, California 94720}                      
\centerline{$^{23}$University of Maryland, College Park, Maryland 20742}        
\centerline{$^{24}$University of Michigan, Ann Arbor, Michigan 48109}           
\centerline{$^{25}$Michigan State University, East Lansing, Michigan 48824}     
\centerline{$^{26}$Moscow State University, Moscow, Russia}                     
\centerline{$^{27}$University of Nebraska, Lincoln, Nebraska 68588}             
\centerline{$^{28}$New York University, New York, New York 10003}               
\centerline{$^{29}$Northeastern University, Boston, Massachusetts 02115}        
\centerline{$^{30}$Northern Illinois University, DeKalb, Illinois 60115}        
\centerline{$^{31}$Northwestern University, Evanston, Illinois 60208}           
\centerline{$^{32}$University of Notre Dame, Notre Dame, Indiana 46556}         
\centerline{$^{33}$University of Oklahoma, Norman, Oklahoma 73019}              
\centerline{$^{34}$University of Panjab, Chandigarh 16-00-14, India}            
\centerline{$^{35}$Institute for High Energy Physics, 142-284 Protvino, Russia} 
\centerline{$^{36}$Purdue University, West Lafayette, Indiana 47907}            
\centerline{$^{37}$Rice University, Houston, Texas 77251}                       
\centerline{$^{38}$Universidade Estadual do Rio de Janeiro, Brazil}             
\centerline{$^{39}$University of Rochester, Rochester, New York 14627}          
\centerline{$^{40}$CEA, DAPNIA/Service de Physique des Particules, CE-SACLAY,   
                   France}                                                      
\centerline{$^{41}$Seoul National University, Seoul, Korea}                     
\centerline{$^{42}$State University of New York, Stony Brook, New York 11794}   
\centerline{$^{43}$SSC Laboratory, Dallas, Texas 75237}                         
\centerline{$^{44}$Tata Institute of Fundamental Research,                      
                   Colaba, Bombay 400005, India}                                
\centerline{$^{45}$University of Texas, Arlington, Texas 76019}                 
\centerline{$^{46}$Texas A\&M University, College Station, Texas 77843}         
}                                                                               


\maketitle

\vspace{-0.2cm}
\begin{abstract}
A measurement of the cross section for 
production of single, isolated photons is reported for 
transverse energies in the range of 10--125~GeV, 
for two regions of pseudorapidity, $|\eta|<0.9$
and $1.6<|\eta|<2.5$. 
The data represent 12.9~pb$^{-1}$ of integrated luminosity 
accumulated in $\bar{p}p$ collisions at $\sqrt{s} = 1.8\,$TeV 
and recorded with the \D0\ detector at the Fermilab Tevatron Collider.
The background, predominantly from jets which fragment to neutral
mesons, is estimated using the longitudinal shower shape in
the calorimeter. 
In both pseudorapidity regions the cross section
is found to agree with the next-to-leading order
QCD prediction for $30 \simle\etg\simle 80\,$GeV.
\end{abstract}

\vspace{-1.cm}

\newpage

\begin{figure}[t]
\vbox to 12cm{
\vfill
\includegraphics{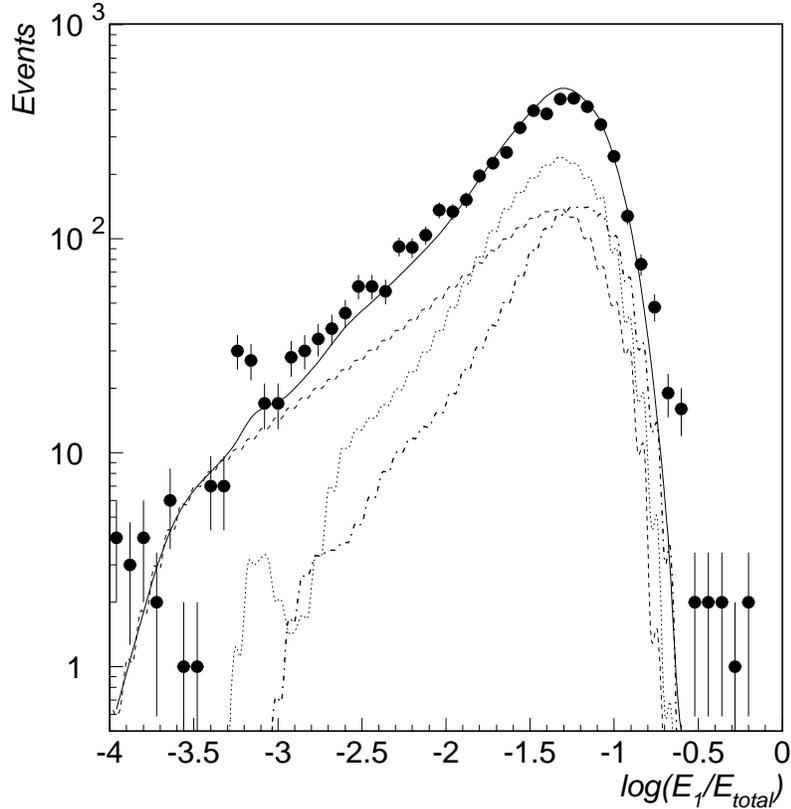}
\vfill
}
\caption{Distribution of $\log{(E_1/E_{\rm total})}$ for 
$\etg=40\pm 5\,{\rm GeV}$ central photon candidates (solid points),
and the fitted distribution (solid curve) made up of Monte Carlo
photons (dashes), 
neutral pions (dots), and $\eta$ mesons (dot-dash curve).
The neutral pion and $\eta$ meson distributions at small
$\log{(E_1/E_{\rm total})}$ fluctuate due to limited
Monte Carlo statistics.}  
\label{emf}
\end{figure}

Direct photon production probes the parton-parton interaction without 
the ambiguities associated with jet identification,
fragmentation and energy measurement.  
In $p \overline p$ collisions at $\sqrt{s} = 1.8\,$TeV,
the dominant mode of production for
low transverse energy photons is
through gluon Compton scattering.  The cross section is thus
sensitive to the gluon distribution in the proton (and antiproton) 
at low momentum fractions $x$.  
With a photon of transverse energy $\etg \geq 10\,$GeV and 
pseudorapidity $\eta \leq 2.5$ ($\eta = - {\rm ln\,tan}{\theta \over 2}$, where
$\theta$ is the polar angle with respect to the proton beam), gluons
with momentum fractions as low as $x \sim 10^{-3}$ contribute to the cross
section.  

Previous collider experiments,
including UA2\cite{uatwo} and CDF\cite{cdf},
have reported an excess of photons at low
$\etg (\simle 30\,{\rm GeV})$ compared with the 
next-to-leading order (NLO) QCD prediction.  This may be explained by 
additional gluon radiation beyond that included in the NLO 
calculation\cite{kt}, or by modified parton
distributions and fragmentation contributions\cite{vvv}.

This letter presents the first measurement\cite{fahey} of 
the cross section for production of isolated photons  
in $p\overline p$ collisions at $\sqrt{s} = 1.8\,$TeV
with pseudorapidity $1.6 < |\eta|< 2.5$, and 
a new measurement\cite{liu} of the isolated photon cross
section in the central region ($|\eta|<0.9$).

Photons are identified in the \D0\ detector\cite{dzero} using a
uranium/liquid argon sampling calorimeter housed in 
a central and two forward cryostats.  The
calorimeters cover
the region of $|\eta| \simle 4$ and have electromagnetic energy resolution
$\sigma_E/{E} = 15\%/\sqrt{E (\rm GeV)} \oplus 0.3\%$.
In both central and forward regions the electromagnetic section
is segmented into 
four longitudinal layers (EM1--EM4) of 2, 2, 7, and
10 radiation lengths respectively; the transverse segmentation is into towers 
of size (in pseudorapidity and azimuthal angle)
$\Delta \eta \times \Delta \phi = 0.1 \times 0.1$ ($0.05 \times 0.05$ at shower
maximum in EM3).
The central and forward drift chambers in front of the calorimeter allow 
photons to be distinguished from electrons and photon conversions 
by ionization measurement.  

The data presented here represent $12.9\pm 0.7\,{\rm pb}^{-1}$ of integrated
luminosity recorded during 1992--93. 
The detector used a three-level triggering system.
The first level used scintillation counters near the 
beam pipe to detect an inelastic $p\bar{p}$ interaction.  
The second level was a hardware trigger which summed
the electromagnetic energy in calorimeter towers of size
$\Delta \eta \times \Delta \phi =0.2 \times 0.2 $. 
The data used in this analysis were taken with single tower
energy thresholds of 2.5, 7, and 10 GeV; all except the highest
threshold were prescaled.  The third level was a software 
trigger in which clusters of calorimeter cells were formed and
loose cuts made on shower shape. 
The cluster energy thresholds used at this level were 6, 14, and 30~GeV
respectively.

Photon candidates were selected as follows.  
Fiducial cuts were applied to select candidate clusters away from the
calorimeter edges:
the clusters were restricted to the regions $| \eta | < 0.9 $ and
$1.6 < |\eta|< 2.5$, and in the central region 
were required to be more than 1.6~cm from the azimuthal calorimeter
module boundaries.
Events where the vertex was more than 50~cm from its nominal position
were discarded.  The resulting acceptance is $A=0.73 \pm 0.01$
($0.86 \pm 0.01$) in the central (forward) regions.

\begin{figure}[t]
\vbox to 12cm{
\vfill
\includegraphics{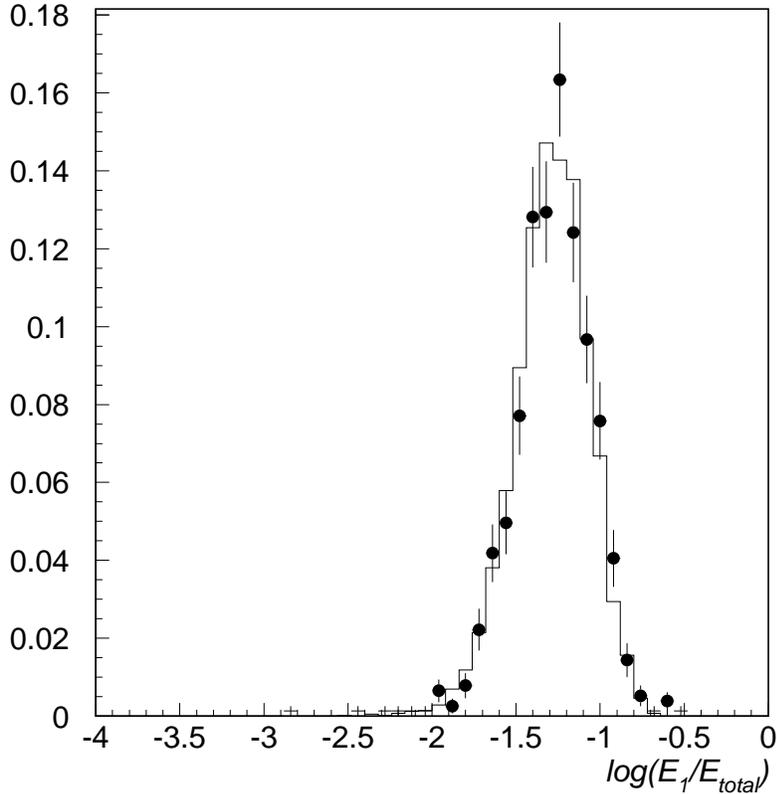}
\vfill
}
\caption{Normalized distributions of $\log{(E_1/E_{\rm total})}$ for 
central electrons from $W\rightarrow e\nu$ events (data points),
and for single Monte Carlo electrons with $E_T = 40\,$GeV (histogram).}
\label{wemf}
\end{figure}

The remaining clusters were identified as photon candidates.  
No drift chamber tracks were allowed in a tracking road
($\Delta\theta \times\Delta\phi
\approx 0.2\times 0.2$ radians) between the calorimeter cluster and
the primary vertex.  
The efficiency of this requirement was 
estimated to be $0.85\pm 0.01~(0.61\pm 0.03)$ in the
central (forward) regions.  (The inefficiency is due to photon conversions 
and overlaps with charged tracks from the underlying event.)
The photon candidate shower 
was required to have a shape consistent with that of a single electromagnetic
shower and 
to have more than 96\% of its energy in the electromagnetic section
of the calorimeter.
The candidates were
required to be isolated by a cut on the transverse energy in the
annular region between
${\cal R}=\sqrt{\Delta\eta^2 + \Delta\phi^2}=0.2$ and ${\cal R}=0.4$ 
around the cluster:
$E_T^{{\cal R}\leq 0.4} - E_T^{{\cal R}\leq 0.2} < 2\,$GeV. 
Finally, the missing transverse energy of the event was 
required to be less than 20~GeV
to reject electrons from 
$W\to e\nu$ decays and events with large amounts of calorimeter noise.
The efficiency of these last three cuts was estimated as
a function of $\etg$ using 
a detailed Monte Carlo simulation of the detector and verified using
$Z \to ee$ events.  The value obtained was $0.92 \pm 0.03~(0.77\pm0.06)$
at $\etg=40\,{\rm GeV}$ for central (forward) photons.  

\begin{figure}[t]
\vbox to 12cm{
\vfill
\includegraphics{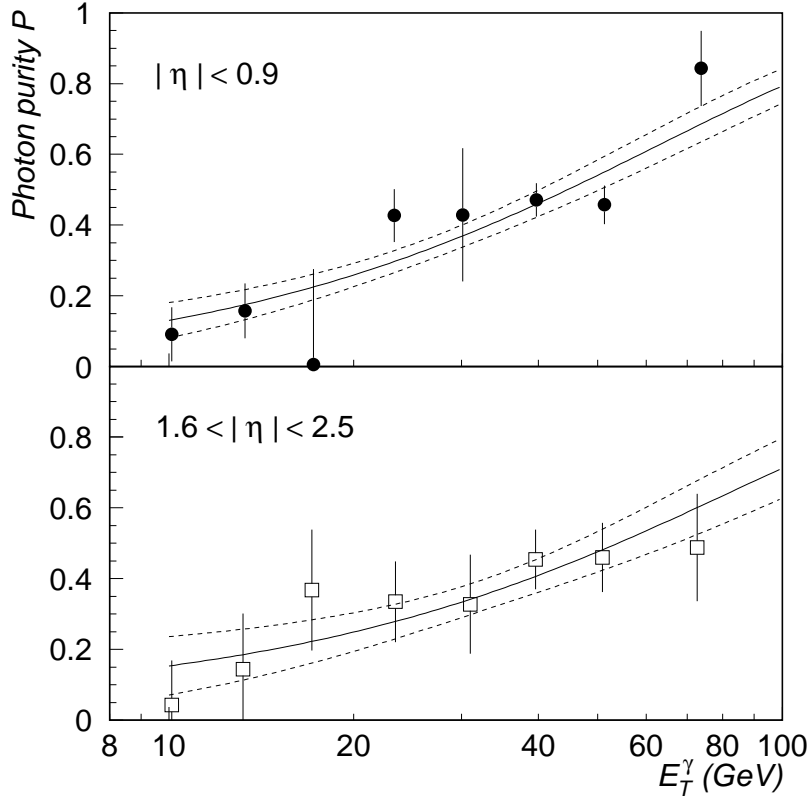}
\vfill
}
\caption{Efficiency-corrected photon purity
(${\cal P}$) vs $E_{T}^{\gamma}$ for central and forward photons.
The solid lines show 
fits of the form $1-e^{-(a+b E_T)}$ and the dashed lines
indicate the range of uncertainty thereon.}
\label{pur}
\end{figure}

The primary experimental challenge in the measurement of the direct photon 
cross section is the extraction of the prompt photon signal from the copious 
backgrounds due to $\pi^0$ and $\eta$ mesons produced in jets which
subsequently
decay to photons.  While the bulk of this jet background is rejected by the 
selection criteria listed above, especially by the requirement that
the photon candidates be isolated, substantial contamination remains.
This comes predominantly from fluctuations in the jet fragmentation which lead 
to a single meson carrying most of the jet energy.  
If the meson has transverse energy above about
$10\,$GeV, the showers from its two decay photons
coalesce and mimic a single photon shower in the calorimeter.  

The fraction of candidates fulfilling the selection
criteria which are genuine direct photons (the purity ${\cal P}$) was 
determined using the energy $E_1$ deposited in the first layer (EM1) of the
calorimeter.  Neutral meson decays produce two photons, and so the
probability that at least one of them undergoes a conversion to an
$e^+e^-$ pair in the calorimeter cryostat and first absorber plate
is roughly twice that for a single photon. Meson
showers therefore start earlier than photon showers leading to larger $E_1$.  
A typical distribution of
$\log{(E_1/E_{\rm total})}$ is shown in Fig.~\ref{emf}.  
The distribution is fit as the sum of a 
photon signal plus $\pi^0$ and $\eta$ meson backgrounds.  
Fitting was done separately for the central and forward samples
for each $\etg$ bin,
using $\chi^2$ minimization, and constraining the fractions of signal and
background to lie in the range $[0,1]$. The results presented use 
a production ratio of $\eta/\pi^0 = 1.0$\cite{etapiratio}, 
but all values between 0.50 and
1.25 give essentially indistinguishable results for ${\cal P}$ (since
the distributions of $\log{(E_1/E_{\rm total})}$ for
$\pi^0$ and $\eta$ mesons are similar).
The Monte Carlo calculation combines a detailed simulation of the
calorimeter with overlaid minimum bias events from data 
to model noise, pileup and the underlying event. Its ability to correctly 
model the $E_1/E_{\rm total}$ distribution has been verified using samples of
electrons from $W\to e\nu$ events taken with the same trigger requirements,
as shown in Fig.~\ref{wemf}.

\begin{figure}[tb]
\vbox to 12cm{
\vfill
\includegraphics{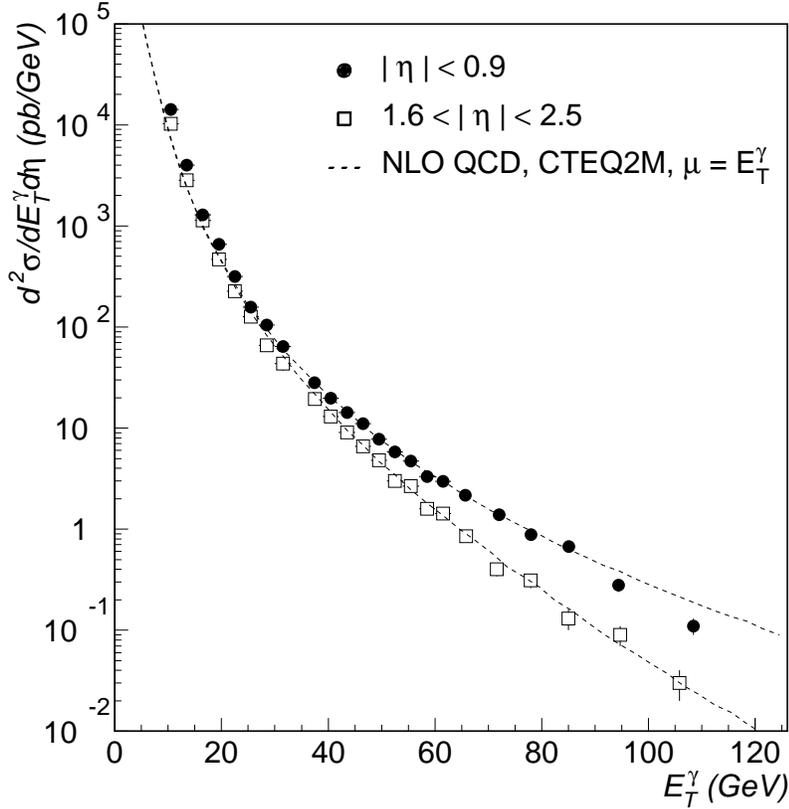}
\vfill
}
\caption{
The inclusive isolated photon cross section 
$\sigma_e = d^2\sigma/dE_T^\gamma\,d\eta$
as a function of photon transverse energy $\etg$, for central (circles)
and forward regions (squares).  The errors are statistical only.  The 
NLO QCD calculated cross sections $\sigma_t$, using CTEQ2M parton 
distributions with $\mu=\etg$, are shown for comparison.}
\label{log}
\end{figure}

\begin{figure}[htb]
\vbox to 12cm{
\vfill
\includegraphics{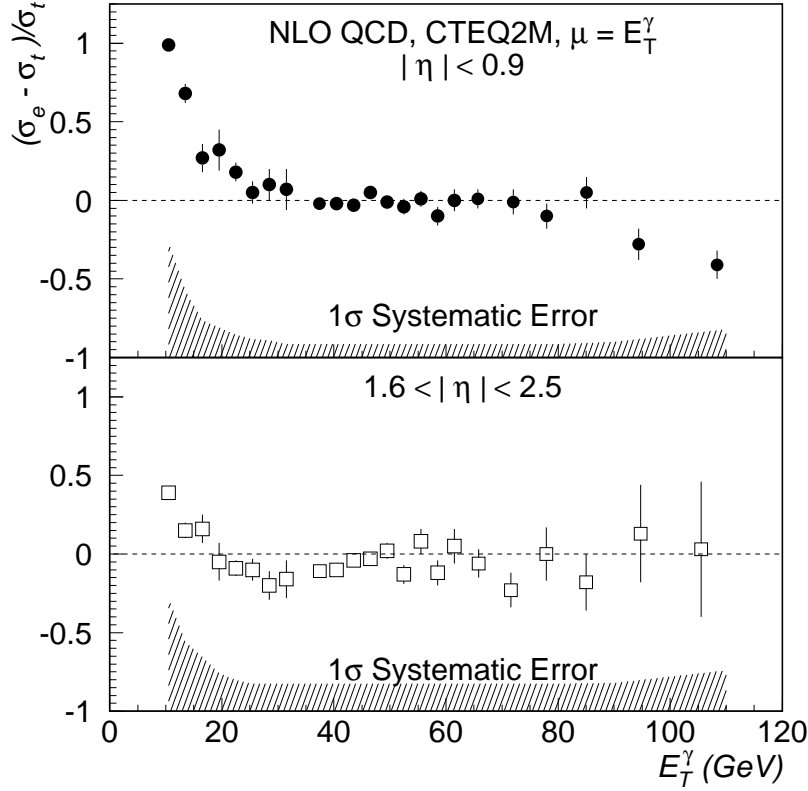}
\vfill
}
\caption{Difference between the measured isolated photon cross section
$\sigma_e$ and the NLO QCD prediction $\sigma_t$, normalized to the latter.
The shaded bands show the magnitude of the 
combined systematic errors ($1\,\sigma$) for each of the two regions.}
\label{lin}
\end{figure}

\begin{figure}[htb]
\vbox to 12cm{
\vfill
\includegraphics{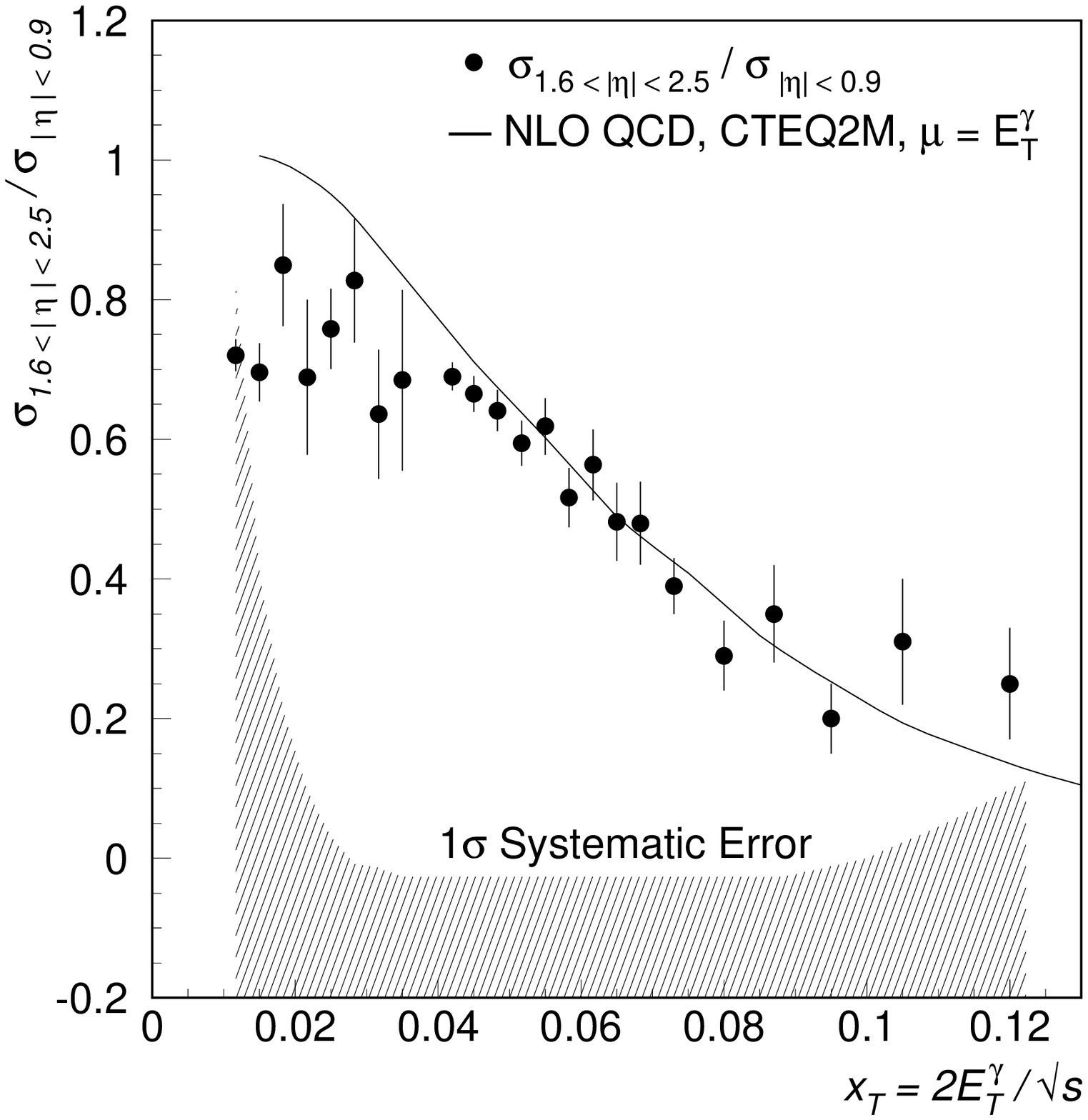}
\vfill
}
\caption{Ratio of inclusive isolated photon cross sections
$d^2\sigma/dx_T\,d\eta$ in the forward region to the central region.
The NLO QCD prediction with CTEQ2M parton 
distributions is shown for comparison. The shaded band shows the 
magnitude of the combined systematic error.}
\label{gluon}
\end{figure}

The combined statistical and systematic error on the purity ${\cal P}$,
at each $\etg$ point ($\etg = 10$, 13,  17,  23,  30,  40,  
51, and  74~GeV), was estimated 
by inflating by $\sqrt{\chi^2}$
the error given by the fit to $\log(E_1/E_{\rm total})$
for that $\etg$.  The factor of $\sqrt{\chi^2}$ accounts for systematic
differences between the Monte Carlo distributions and the data.
It was typically 1.3 in the central region and 1.6 in the
forward region. 
The central and forward photon purities were then corrected by the 
$\etg$-dependent efficiencies and fit to the form:
\begin{eqnarray}
   {\cal P} = 1 - e^{-(a+b E_T^{\gamma})}.
\end{eqnarray}
The data points, fits, and fit errors for ${\cal P}$ 
are shown in Fig.~\ref{pur}.  

A second method of purity estimation 
was used to check the results from
the calorimeter energy deposition method.
It also takes advantage of the difference in
conversion probability between single photons and
background.
In this case the material between the interaction
point and the central or forward drift chamber (CDC or FDC) is considered as 
a converter and 
conversions are tagged as tracks with twice minimum ionizing energy using 
the ionization measurement in the CDC or FDC. 
The results from the two methods are found
to be consistent.  The conversion method has larger statistical
errors, since only $\sim 10$\% of photons convert, and therefore
it was not used in the fit.

The differential cross section $d^2\sigma/d\etg\,d\eta$ is plotted as a 
function of $\etg$ in 
Fig.~\ref{log}.  
The next-to-leading order QCD calculation was generated using 
a program due to Baer, Ohnemus, and Owens\cite{owens} which includes 
$\gamma+{\rm jet}$,
$\gamma+{\rm dijet}$ and dijet$+$bremsstrahlung final states.  This last
is generated from dijet states with collinear bremsstrahlung
obtained from a phenomenological photon fragmentation function.  In this case,
a third collinear jet was created with the remaining jet energy, so that 
the isolation cut could be modeled including jet energy fluctuations.  
In all cases the parton energies were smeared using the measured D\O\ 
electromagnetic and jet resolutions.  
The isolation cut is imposed by computing the distance ${\cal R}$
in $\eta$-$\phi$ space
between the photon and any of the jets, and then rejecting events 
with a smeared jet $E_T > 2\,{\rm GeV}$ within ${\cal R}\leq0.4$ of 
the photon.  
(Use of smeared photon and jet energies changes the QCD prediction by
less than 4\% but better represents the actual measurement.)

The CTEQ2M 
parton distributions\cite{CTEQ} and renormalization scale
$\mu=\etg$ were used.  If instead the
CTEQ2MF or CTEQ2MS parton distributions were used, or scales of 
$\mu=2\etg$ or $\etg/2$ were employed,
then the predicted cross section changes by $\simle 6$~\%. 

Figure~\ref{lin}
shows a plot of $(\sigma_e - \sigma_t)/\sigma_t$ where
$\sigma_e$ and $\sigma_t$ are respectively the experimental and 
theoretical values of the differential
cross section $d^2\sigma/d\etg\,d\eta$.
The shaded band in the figure shows the magnitude of
the systematic error, which is estimated by adding in quadrature the
uncertainties resulting from the acceptance ($\sim 1$\%), the trigger 
and selection efficiencies (3--5\%), 
the photon purity (Fig.~3), the luminosity (5\%),
and the electromagnetic energy scale of the calorimeter 
(1\% in the central, 4\% in the forward region).
The measured cross sections are 
in good agreement with the NLO QCD prediction for both central
and forward regions for moderate transverse energies,
$30 \simle \etg \simle 80\,{\rm GeV}$.
The data points for both the central and forward cross sections 
lie above the NLO QCD prediction at lower transverse 
energies, but given the magnitude of the systematic error no strong conclusion
can be drawn. 
At the highest transverse energies probed, the data for the central region 
lie below the QCD prediction.  Above $\etg = 74\,$GeV 
the photon purity $\cal{P}$ is the result of an extrapolation;
the increased systematic error at large $\etg$ reflects the resulting
uncertainty. 

The ratio of forward to central cross sections is shown in Fig.~\ref{gluon}.
The systematic error on the ratio is estimated by adding those for the
two regions in quadrature, with the exception of the luminosity 
uncertainty which cancels.
The ratio is in good agreement with the NLO QCD expectation for large 
transverse momenta, but the data lie below NLO QCD for 
$x_T \simle 0.04$ ($\etg \simle 36\,{\rm GeV}$).
Given the magnitude of the systematic error, no strong conclusion can be
drawn.  
This ratio of cross sections, which depends on our unique measurement of the
forward photon cross section, could be used to constrain the gluon distribution
function at low $x$.  However, a complete understanding of the origin 
of the low-$\etg$ behavior of the photon cross section is needed before 
information on the gluon distribution can be extracted.

\section*{Acknowledgements}

%
We thank the Fermilab Accelerator, Computing, and Research Divisions, and
the support staffs at the collaborating institutions for their contributions
to the success of this work.   We also acknowledge the support of
the U.S. Department of Energy,
the U.S. National Science Foundation,
the Commissariat \`a L'Energie Atomique in France,
the Ministry for Atomic Energy and the Ministry of Science and Technology 
Policy in Russia,
CNPq in Brazil,
the Departments of Atomic Energy and Science and Education in India,
Colciencias in Colombia, 
CONACyT in Mexico,
the Ministry of Education, Research Foundation and KOSEF in Korea,
CONICET and UBACYT in Argentina,
and the A.P. Sloan Foundation.


\vspace{-0.3cm}

\begin{references}
\vspace{-1.5cm}

%
\bibitem[*]{beijing}
Visitor from IHEP, Beijing, China.

\bibitem[\dag]{ecuador}
Visitor from Univ. San Francisco de Quito, Ecuador.

\vskip 0.25cm


\bibitem{uatwo} 
UA2 Collaboration, J. Alitti {\it et al.,}
Phys. Lett. {\bf B263} (1991) 544.

\bibitem{cdf} 
CDF Collaboration, 
F. Abe {\it et al.,} Phys. Rev. Lett. {\bf 68} (1992) 2734;\\
CDF Collaboration,
F. Abe {\it et al.,} Phys. Rev. D~{\bf 48} (1993) 2998;\\
CDF Collaboration,
F. Abe {\it et al.,} Phys. Rev. Lett. {\bf 73} (1994) 2662.

\bibitem{kt}
J. Huston {\it et al.,} Phys. Rev. D~{\bf 51} (1995) 6139.

\bibitem{vvv} 
M. Gl\"{u}ck, L.E.~Gordon, E.~Reya, and W.~Vogelsang,
Phys. Rev. Lett. {\bf 73} (1994) 388;\\
W. Vogelsang and A. Vogt, Nucl. Phys. {\bf B453} (1995) 334;\\
E. Quack and D.K.~Srivastava, ``A Global Test of QCD Theories for Direct Photon
Production,'' GSI-94-40, August 1995.

\bibitem{fahey} S.T.~Fahey, Ph.D. Thesis, Michigan State University, 1995
(unpublished).  

\bibitem{liu} Y.-C.~Liu, Ph.D. Thesis, Northwestern University, 1996
(unpublished).  
Accessible via the World Wide Web at\\
{\tt http://d0sgi0.fnal.gov/publications\_talks/thesis/thesis.html}.

\bibitem{dzero} D\O\ Collaboration, S. Abachi {\em et al.}, 
Nucl. Instrum. Methods {\bf A338} (1994) 185.

\bibitem{etapiratio} 
CDF Collaboration, 
F. Abe {\it et al.,} Phys. Rev. D~{\bf 48} (1993) 2998. 

\bibitem{owens} H. Baer, J. Ohnemus, and J.F. Owens,
Phys. Rev. D~{\bf 42} (1990) 61.

\bibitem{CTEQ} CTEQ Collaboration, J. Botts {\em et al.},
Phys. Lett. {\bf B304} (1993) 159.

\end{references}
\end{document}